\documentclass[sn-mathphys,Numbered]{sn-jnl}

\usepackage{graphicx}%
\usepackage{multirow}%
\usepackage{amsmath,amssymb,amsfonts}%
\usepackage{amsthm}%
\usepackage{mathrsfs}%
\usepackage[title]{appendix}%
\usepackage{xcolor}%
\usepackage{textcomp}%
\usepackage{manyfoot}%
\usepackage{booktabs}%
\usepackage{algorithm}%
\usepackage{algorithmicx}%
\usepackage{algpseudocode}%
\usepackage{listings}%

\usepackage{physics}

\theoremstyle{thmstyleone}%
\theoremstyle{thmstyletwo}%

\theoremstyle{thmstylethree}%

\begin{document}

\title[Hierarchical phased-array antennas coupled to Al KIDs: a scalable architecture for multi-band mm/submm focal planes]{Hierarchical phased-array antennas coupled to Al KIDs: a scalable architecture for multi-band mm/submm focal planes}

\author*[1]{\sur{Jean-Marc Martin}}\email{jeanmarc@caltech.edu}

\author[1]{\sur{Junhan Kim}}

\author[2]{\sur{Fabien Defrance}}

\author[1]{\sur{Shibo Shu}}

\author[2]{\sur{Andrew D. Beyer}}

\author[2]{\sur{Peter K. Day}}

\author[1]{\sur{Jack Sayers}}

\author[1]{\sur{Sunil R. Golwala}}

\affil[1]{\orgname{California Institute of Technology}, \orgaddress{\street{1200 E. California Blvd.}, \city{Pasadena}, \postcode{91125}, \state{California}, \country{USA}}}

\affil[2]{\orgname{Jet Propulsion Laboratory, California Institute of Technology}, \city{Pasadena}, \postcode{91109}, \state{California}, \country{USA}}



\abstract{We present the optical characterization of 
two-scale hierarchical phased-array antenna kinetic inductance detectors (KIDs) for millimeter/submillimeter wavelengths.  
Our KIDs have a lumped-element architecture with parallel plate capacitors and aluminum inductors.  
The incoming light is received with a hierarchical phased array of slot-dipole antennas, split into 4 frequency bands (between 125~GHz and 365~GHz) with on-chip lumped-element band-pass filters, and routed to different KIDs using microstriplines. 
Individual pixels detect light for the 3 higher frequency bands 
(190--365~GHz)
and the signals from four individual pixels are coherently summed to create a larger pixel detecting light for 
the lowest-frequency band
(125--175~GHz).
The spectral response of the band-pass filters was measured using Fourier transform spectroscopy (FTS), the far-field beam pattern of the phased-array antennas was obtained using an infrared source mounted on a 2-axis translating stage, and the optical efficiency of the KIDs was characterized by observing loads at 294~K and 77~K.  
We report on the results of these three measurements. 
}


\maketitle

\section{Introduction}\label{sec1}

Future 
large-aperture (30~m to 50~m) mm/submm ground based telescopes (e.g., CSST, CMB-HB, AtLAST) will require detectors able to observe several spectral bands within the range 
75~GHz to 415~GHz in order to study cold, dusty sources, mm/submm time-domain sources, the circumgalactic medium and galaxy cluster intracluster medium using the Sunyaev-Zel'dovich effect, and the cosmic microwave background intensity and polarization on sub-arcminute scales.
Given the cost of such large apertures, it is most efficient if each pixel in the focal plane can sense multiple spectral bands.  It is optimal to match the pixel size to the scaling of the diffraction spot size with frequency to avoid oversampling (excessive detector count) or undersampling (loss of angular resolution) the focal plane.  Hierarchical antennas, first proposed in~\cite{Goldin:2002,Goldin:2003} and first demonstrated by \cite{cukierman_hierarchical_2018}, can meet this need by coherently summing signals from individual pixels to provide this frequency scaling of pixel size, as first noted in \cite{Stacey:2006,Ji:2014}.

\begin{figure}
    \centering
    \includegraphics[width=0.9\textwidth]{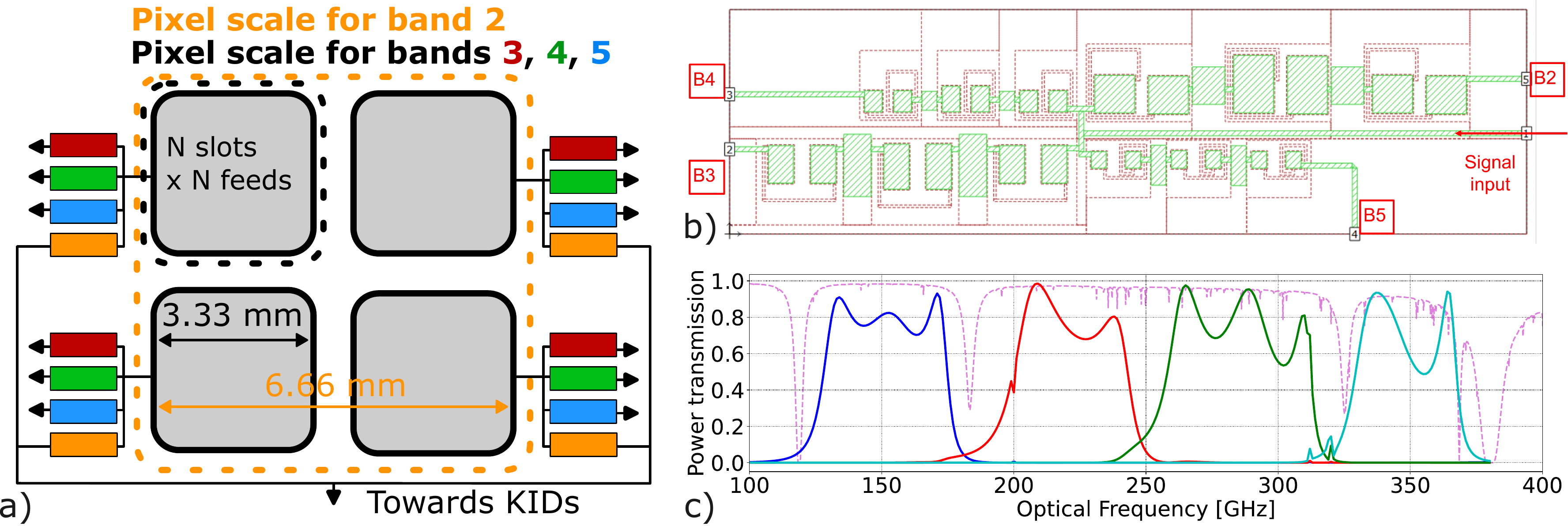}
    \caption{
    a)~Schematic of the hierarchical phased-array antenna.  Each sub-element is similar to the antenna described in \cite{Golwala2012}, though the SiN$_x$ dielectric has been replaced by a-Si:H and the shunt capacitor and microstripline dimensions adjusted for the differing dielectric constant. b)~Schematic of the band-pass filter-bank~\cite{shu_multi-chroic_2022}, with the four bands indicated. c)~Sonnet simulations of the band-pass filter-bank with the atmospheric transmission line (pink) overlaid from ATM model with $\text{PWV}=0.5$ at Llano de Chajnantor Observatory. The bands are B2, B3, B4, and B5 in order of increasing frequency.
    }
    \label{fig:ShiboDesign}
\end{figure}

We first introduced the concept for a three-scale hierarchical phased-array slot-dipole antenna coupled to TiN$_x$ KIDs, covering 75--415~GHz, in \cite{Ji:2014}, and we designed and fabricated a two-scale version with six bands over this frequency range.  We found it difficult to model the observed response of these KIDs because TiN$_x$ deviates from Mattis-Bardeen theory~\cite{MattisBardeen:1958}, so we revised the design to use Al KIDs.  We also discovered issues with our initial antenna and band-pass filter design that led us to reduce scope to four bands for an initial demonstration (without reducing the antenna's intrinsic bandwidth).  We reported on the revised design, including preliminary optical efficiency and noise measurements, in Shu et al. \cite{shu_multi-chroic_2022}.
Figure~\ref{fig:ShiboDesign} shows  the two-scale, four-band design.
The four frequency bands correspond to the bands 2 to 5 of the initial 6-band design and are designated as B2, B3, B4, and B5 in the article. 
The designed mean frequency ($\nu_\text{mean}$) and effective bandwidth ($\Delta \nu_\text{eff.}$) of these bands are listed in  Tab.~\ref{tab:bands}.
The use of amorphous hydrogenated silicon (a-Si:H) with a very low loss tangent (close to $10^{-5}$ at RF frequencies, and expected to be 
$\lesssim 10^{-4}$ at millimeter/submillimeter wavelengths) 
in the microstripline ensures the additional length required for hierarchical summing contributes negligible additional loss.

In this paper, we present the phased-array beam-pattern characterization for the four frequency bands, the spectral response of the filter-bank measured using Fourier Transform Spectroscopy (FTS), and the optical efficiency of the KIDs.  
The beam-maps and FTS measurements were obtained using the GPU-accelerated RF readout developed by Minutolo et al. \cite{minutolo_flexible_2019}.
The devices are operated in a cryostat that provides a 244~mK operating temperature\footnote{$^4$He/$^3$He/$^3$He Chase Cryogenics sorption cooler based in a custom cryostat cooled by a Cryomech PT-415 two-stage pulse-tube cooler} and is configured with optical windows so response can be measured up to 40$^\circ$ off axis.

\section{Beam Maps}\label{sec2}

\begin{figure}
    \centering
    \includegraphics[width=0.89\textwidth]{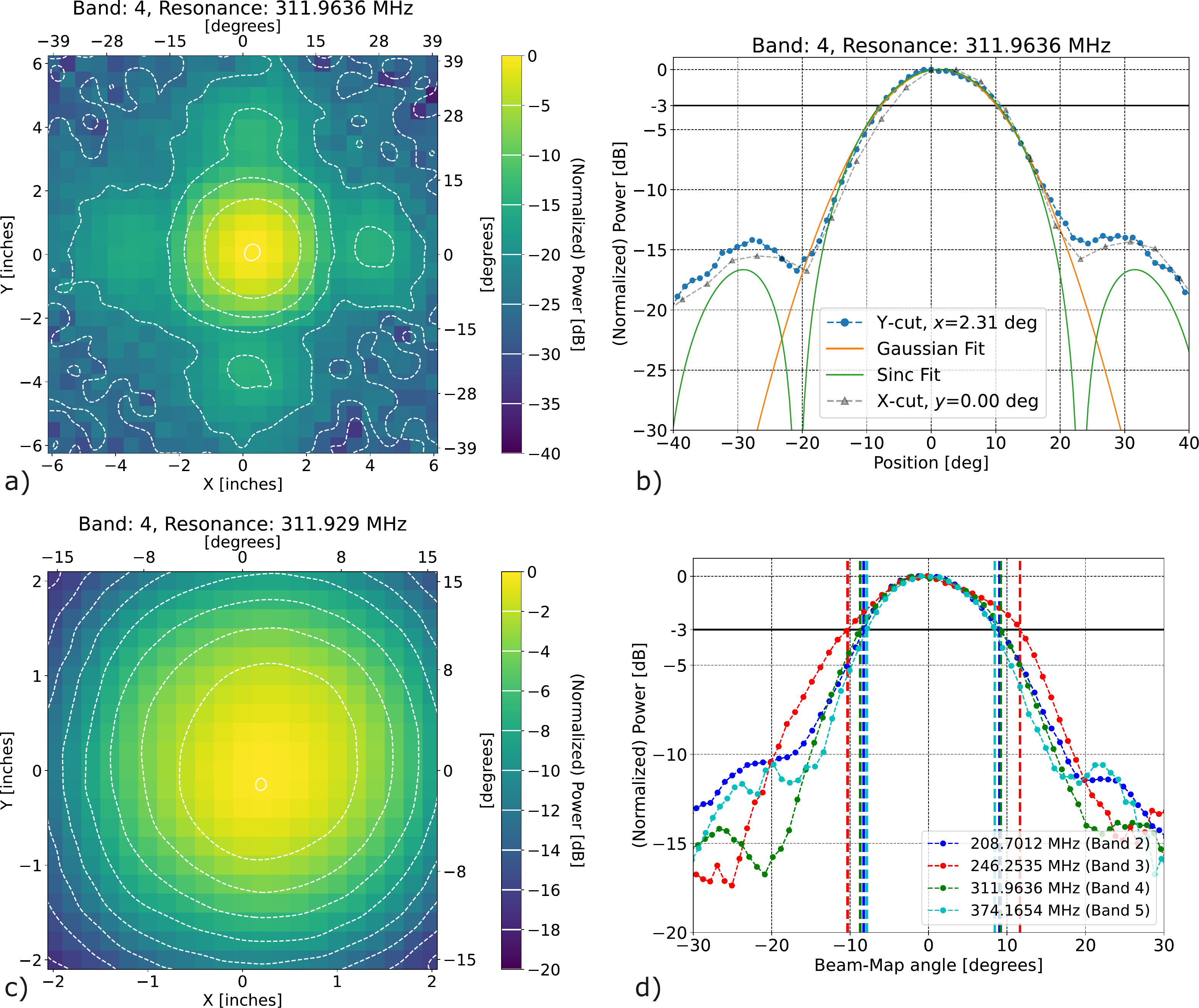}
    \caption{
    a)~Typical two-dimensional beam map for B4.  b)~B4 beam map cross sections at $x=0$ and $y=0$. c)~Close-up on central lobe for B4 beam map.  d) Beam cross sections for all bands.  The vertical lines indicate the FWHMs, reported in Table~\ref{tab:bands}.  The similarity of the B2 and B4 beams demonstrate the hierarchical summing is functional.}
    \label{fig:BeamMap}
\end{figure}

To measure the beam profile of the hierarchical phased-array antenna pixels in each band, we measured the response of the KIDs while scanning 
a blackbody source, chopped at 8~Hz, mounted on a motorized 2-axis linear translation stage at $\simeq 189~\text{mm}$ distance from the KIDs.  The 300~mm travel of the beam-mapper enables measurement of the first beam sidelobe.
We measure the amplitude of the resulting modulation in the RF network transmission $S_{21}(f_{res})$, which is proportional to the KID frequency shift and thus the intensity of light received.
Figure~\ref{fig:BeamMap} shows example beam maps.  Because the antenna illumination pattern in the focal plane is approximately uniform over a square 3.3~mm on a side, we expect the beams to be sinc functions.  Fig.~\ref{fig:BeamMap} compares cross-sectional cuts to the expected beam shape and a phenomenological Gaussian fit to the main lobe.  The model reproduces qualitatively the expected side lobe.  Most importantly, B2's FWHM and sidelobe positions match the expectation from hierarchical summing (pixel size twice as large) and are narrower/closer in than B3's.  Table~\ref{tab:bands} shows the measured FWHMs.  Further modeling will be done to better understand the discrepancy between observed and expected sidelobe positions, the filling-in of the sinc function's first null, and the somewhat elevated shoulder seen in B2's beam.

\section{Band-pass Measurements}\label{sec3}

We use a Martin-Puplett interferometer to do Fourier transform spectroscopy to measure  spectral response.  The interferometer is fed by a $1050~C$ cavity blackbody, chopped at 8 Hz.  We again monitor the KIDs' modulated response.
The resulting interferogram is Fourier-transformed to generate the 
power transmission spectrum/response of the whole system (cryostat window, optical filters, hierarchical antenna, band-pass filters, microstripline, and KIDs) between 100~GHz and 400~GHz.

\begin{figure}
    \centering

    \includegraphics[width=0.9\textwidth]{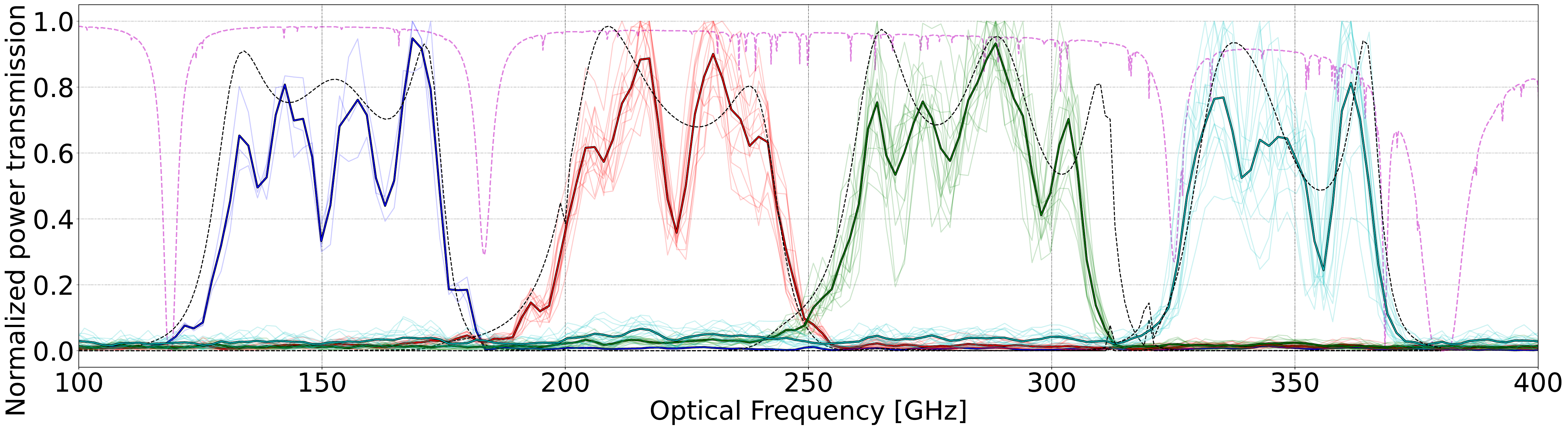}
    \caption{
    Measured power transmission spectrum.  The thin solid lines show the spectral responses of individual KIDs, while the thick solid lines show the mean response for each band.  The dashed black lines show the same Sonnet simulations and the dashed pink line the same atmospheric transmission as in Figure~\ref{fig:ShiboDesign}.}
    \label{fig:FTS}
\end{figure}

Figure~\ref{fig:FTS} shows the measured spectral responses, both for each KID and averaged over each band, along with the Sonnet simulation of the band-pass filters. (We expect the response of all other elements to be smooth functions of frequency, nearly flat over any individual band.)
Table~\ref{tab:bands} show the effective bandwidth and band center of each band, obtained via $\Delta\nu_\text{eff.} = \int t_\text{norm}(\nu)\, \dd \nu$ and $\nu_\text{mean}=\int t_\text{norm}(\nu)\,\nu\, \dd \nu / \Delta\nu_\text{eff.}$, with $t_\text{norm}(\nu) $ the normalized power transmission.  The results differ from the Sonnet simulation by 14-20\% for $\Delta\nu_\text{eff.}$ and 1-5\% for $\nu_\text{mean}$, where the former can be attributed primarily to in-band transmission variation rather than the position of the band edges (with the exception of B4).
The internal minima in the bands are yet to be investigated on if they are due to the free-space optical path or from the transmission lines.

\section{Optical efficiency}\label{sec4}

\begin{figure}
    \centering
    \includegraphics[width=0.95\textwidth]{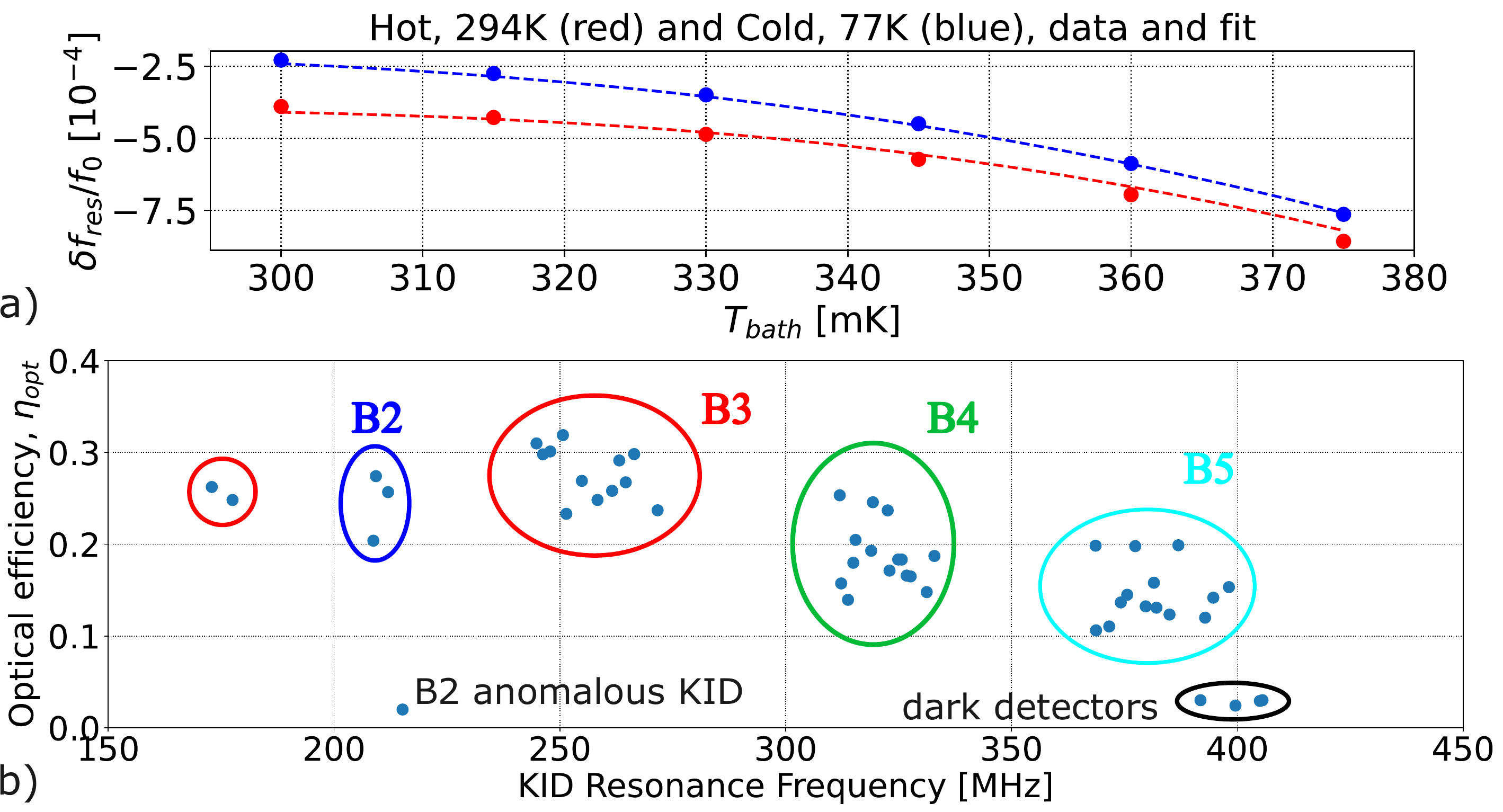}
    \caption{a) Data points (dots) and corresponding fits (dashed lines) of $\delta f_{res} / f_0$ variation as a function of $T_\text{bath}$ for $T_\text{load} = 77$~K (blue) and $T_\text{load} = 294$~K (red). b) Fitted optical efficiency ($\eta_{opt}$) for each KID. Colored circles indicate the spectral band of each KID. The anomalous KID is probably not well coupled to its antenna due to a fabrication flaw.  The ``dark detectors'' are intentionally not coupled to antennas.}
    \label{fig:HotCold}
\end{figure}

To measure the end-to-end optical efficiency, we place a warm ($T_{\text{load}} = 294$~K) or cold (($T_{\text{load}} = 77$~K) blackbody load\footnote{61~cm $\times$ 61~cm panel of WAVASORB\textsuperscript{\textregistered} VHP, immersed in liquid nitrogen for the cold load} in front of the vacuum window, terminating the entire beam that exits the cryostat, and measure the change in KID resonant frequency, repeating the measurement at bath temperatures $T_\text{bath}$ between 300 and 375~mK to break model degeneracies.
A super air knife (Exair) generates a laminar air flow parallel to the cryostat window to prevent water vapor from condensing on the window while observing the 77~K blackbody.
We fit the data to Equation~\ref{eq:fit} (e.g.,~\cite{siegel_multiwavelength_2015}) to measure $\eta_{opt}$ for each KID:

\noindent
\footnotesize
\begin{align}
\frac{\delta f_\text{res}}{f_0} = -\frac{\alpha}{2} \kappa_2\left(\left[\frac{\eta_{opt} \eta_{pb} k_B \Delta\nu_\text{eff.} \qty(T_{\text{load}}+T_\text{exc})}{R V \Delta_0} + n_{\text{th}}^2 + \frac{1}{R\tau_\text{max}} \qty(n_{\text{th}}+\frac{1}{4R\tau_\text{max}})\right]^{1/2} -\frac{1}{2R\tau_\text{max}}\right)
\label{eq:fit}
\end{align}

\normalsize
\noindent where $\delta f_\text{res} = f_\text{res}(T_\text{bath},T_\text{load})-f_0$, $f_0=f_\text{res}(T_\text{bath}=0,T_\text{load}=0)$, $\alpha=0.24$--0.33 is the kinetic inductance fraction, $\eta_{opt}$ is the total optical efficiency, $\eta_{pb}$ is the pair breaking efficiency given in Table~\ref{tab:bands}, 
$T_\text{exc}$ accounts for a fixed additional (``excess'') optical load due to cryostat emission,
$V=3224~\mu\text{m}^3$ is the volume of the KIDs' inductor, $\Delta_0\simeq 206~\mu\text{eV}$ is the gap energy at 0~K, $n_\text{th}$ is the thermal quasi-particle density, $\tau_\text{max}=400~\mu\text{s}$ is the maximum quasi-particle life time \cite{kozorezov_inelastic_2008,barends_quasiparticle_2008,barends_enhancement_2009},  $R=2\Delta_0^2/\qty[N_0\tau_0(k_B T_c)^3]$ is the recombination rate per unit density of quasi-particles, $N_0=1.07\times10^{29}~\text{J}^{-1}\mu\text{m}^{-3}$ is the single-spin electron density of states at the Fermi energy level \cite{siegel_multiwavelength_2015}, $k_B$ is the Boltzmann constant, $T_c\simeq 1.36~\text{K}$ is aluminum critical temperature, $\tau_0=438~\text{ns}$ is the characteristic electron-phonon interaction time \cite{de_visser_quasiparticle_2014,guruswamy_quasiparticle_2014},
and $\kappa_2$ is defined as:

\footnotesize
\begin{equation}
\kappa_2 = \frac{1}{\pi N_0 \Delta_0} \qty[1+\sqrt{\frac{2\Delta_0}{\pi k_b T_\text{bath}}}\exp(-\xi)I_0(-\xi)],
\end{equation}

\normalsize \noindent
with $I_0$ the $0^\text{th}$ order modified Bessel function of the first kind, $\xi=\hbar\omega/(2k_BT_\text{bath})$, $\hbar$ the reduced Planck constant, and $\omega = 2 \pi f_{res}$.
The values of $\alpha$ and $\Delta_0$ used here were obtained for each KID during a previous experiment, we assume $T_c \approx \Delta_0/(1.76 k_B)$, and for each frequency band we use the value of $\Delta\nu_\text{eff.}$ obtained with the FTS measurement.  
We measured $\alpha$ and $\Delta_0$ previously with no optical load on the devices, we assume $T_c \approx \Delta_0/(1.76 k_B)$, and we use the measured values of $\Delta\nu_\text{eff.}$ from Table~\ref{tab:bands}.
In addition to $\eta_{opt}$, we also fit for $f_0$ and $T_\text{exc}$.  Figure~\ref{fig:HotCold} shows a typical dataset and fit and the inferred values of $\eta_{opt}$.  The efficiencies are reasonable considering that they include cryostat optical transmission and that the antireflection layer used for the device substrate is tuned for 200--300~GHz.  More detailed modeling is underway, as are measurements of the nature of the response of the ``dark detectors'' (KIDs not connected to antennas).

\section{Conclusion}\label{sec5}

\begin{table}
\centering
\caption{
Simulated and measured parameters for the four spectral bands: $\nu_\text{mean}$ is the center frequency of the band, $\Delta\nu_\text{eff.}$ is the effective bandwidth, $\eta_\text{pb}$ the pair-breaking efficiency estimated with $T_c=1.36~\text{K}$ and $\nu_\text{mean}$ measured \cite{guruswamy_quasiparticle_2014}.  FWHM is the full width at half maximum of the hierarchical phase-array antenna central lobe, as measured in Fig.~\ref{fig:BeamMap}d and theoretically calculated by $(c / \nu_\text{mean})/ L_\text{antenna}$, where $c$ is the speed of light in vacuum and $L_\text{antenna}$ the size of hierarchical phase-array antenna pixel (6.66~mm for B2 and 3.33~mm for B3, B4, B5).}
\begin{tabular}{|c|c||c|c|c||c|c|c|}
\hline
\multicolumn{2}{|c||}{} & \multicolumn{3}{c||}{Expectation} & \multicolumn{3}{c|}{Measurement} \\
\hline
band & $\eta_{\text{pb}}$ & $\nu_\text{mean}$ & $\Delta\nu_\text{eff.}$ & FWHM & $\nu_\text{mean}$  & $\Delta\nu_\text{eff.}$   & FWHM \\
\hline
2 & 0.64 & 151.02~\text{GHz} & 40.26~\text{GHz} & $17.1^\circ$ & 157.84~\text{GHz} & 34.11~\text{GHz} & $17.3^\circ$ \\
\hline
3 & 0.49 & 219.52~\text{GHz} & 38.59~\text{GHz} & $23.5^\circ$ & 224.16~\text{GHz} & 38.10~\text{GHz} & $22.0^\circ$ \\
\hline 
4 & 0.45 & 282.84~\text{GHz} & 45.60~\text{GHz} & $18.2^\circ$ & 280.33~\text{GHz} & 36.33~\text{GHz} & $17.9^\circ$ \\
\hline 
5 & 0.41 & 346.67~\text{GHz} & 31.12~\text{GHz} & $14.9^\circ$ & 345.12~\text{GHz} & 25.96~\text{GHz} & $16.3^\circ$ \\
\hline
\end{tabular}
\label{tab:bands}
\end{table}


We have demonstrated a two-scale hierarchical antenna/band-pass filter-bank architecture that has beam patterns and spectral bandpasses in reasonable agreement with expectations and good optical efficiency.  Further modeling, to be presented in a future publication, will endeavor to explain deviations from simple sinc function expectations.  We are currently revising the band-pass filter-bank design to reduce ripple, add B1 and B6, and integrate it with a three-scale hierarchical antenna.  Hierarchical antennas are thus a promising technology for maximizing the use of the focal plane of expensive, large-aperture (30--50~m) mm/submm telescopes.

\bmhead{Acknowledgments}
This work has been supported by the JPL Research and Technology Development Fund, the National Aeronautics and Space Administration under awards 80NSSC18K0385 and 80NSSC22K1556, and the Department of Energy Office of High-Energy Physics Advanced Detector Research program under award DE-SC0018126.
The research was carried out in part at the Jet Propulsion Laboratory, California Institute of Technology, under a contract with the National Aeronautics and Space Administration (80NM0018D0004).

\bigskip





\bibliography{LTD20}

\end{document}